\documentclass[twocolumn,letter]{aastex631}  
\usepackage{amsmath}	
\usepackage{graphicx}
\usepackage{txfonts}

\newcommand{\msun}{{$M_{\odot}$}}

\newcommand{\pers}{s$^{-1}$ }
 
\newcommand{\gtsima}{$\; \buildrel > \over \sim \;$}
\newcommand{\ltsima}{$\; \buildrel < \over \sim \;$}
\newcommand{\prosima}{$\; \buildrel \propto \over \sim \;$}
\newcommand{\gsim}{\lower.5ex\hbox{\consistegtsima}}
\newcommand{\lsim}{\lower.5ex\hbox{\ltsima}}
\newcommand{\simgt}{\lower.5ex\hbox{\gtsima}}
\newcommand{\simlt}{\lower.5ex\hbox{\ltsima}}
\newcommand{\simpr}{\lower.5ex\hbox{\prosima}}

\begin{document}  

\title{Evaporation of Close-in Sub-Neptunes by Cooling White Dwarfs}

\author[0000-0001-5802-6041]{Elena Gallo}
\affiliation{
    Department of Astronomy, University of Michigan, 1085 S University, Ann Arbor, MI 48109, USA}
\author[0000-0002-4586-748X]{Andrea Caldiroli}
\affiliation{
    Fakult\"at f\"ur Mathematik, Universit\"at Wien, Oskar-Morgenstern-Platz 1, A-1090 Wien, Austria}
\author[0000-0001-5252-5042]{Riccardo Spinelli}
\affiliation{INAF, Osservatorio Astronomico di Palermo, Piazza del Parlamento 1, 90134 Palermo, Italy}
\author[0000-0001-9113-3906]{Federico Biassoni}
\affiliation{Dipartimento di Scienza e Alta Tecnologia, Universit\`a degli Studi dell'Insubria, via Valleggio 11, I-22100 Como, Italy}
\affiliation{INAF, Osservatorio Astronomico di Brera, Via E. Bianchi 46, I-23807 Merate, Italy}
\author[0000-0003-3291-3704]{Francesco Haardt}
\affiliation{Dipartimento di Scienza e Alta Tecnologia, Universit\`a degli Studi dell'Insubria, via Valleggio 11, I-22100 Como, Italy}
\affiliation{INAF, Osservatorio Astronomico di Brera, Via E. Bianchi 46, I-23807 Merate, Italy}
\affiliation{INFN, Sezione Milano-Bicocca, P.za della Scienza 3, I-20126 Milano, Italy}
\author[0000-0002-9521-9798]{Mary Anne Limbach}
\affiliation{
    Department of Astronomy, University of Michigan, 1085 S University, Ann Arbor, MI 48109, USA}
\author[0000-0001-9256-5508]{Juliette Becker}
\affiliation{Department of Astronomy, University of Wisconsin-Madison, 475 N Charter St, Madison, WI 53706, USA}
\author[0000-0002-8167-1767]{Fred C. Adams}
\affiliation{Department of Physics, University of Michigan, 450 Church Street, Ann Arbor, MI 48109, USA
}
\affiliation{
    Department of Astronomy, University of Michigan, 1085 S University, Ann Arbor, MI 48109, USA}

\begin{abstract}
Motivated by the recent surge in interest concerning white dwarf (WD) planets, this work presents the first numerical exploration of WD-driven atmospheric escape, whereby the high-energy radiation from a hot/young WD can trigger the outflow of the hydrogen-helium envelope for close-in planets. As a pilot investigation, we focus on two specific cases: a gas giant and a sub-Neptune-sized planet, both orbiting a rapidly cooling WD with mass $M_\ast$ = 0.6 \msun\ and separation $a$ = 0.02 AU. In both cases, the ensuing mass outflow rates exceed $10^{14}$ g sec$^{-1}$ for WD temperatures greater than $T_{\rm WD}\simgt 50,000$ K. At $T_{\rm WD}\simeq$ 18,000 K [/22,000 K], the sub-Neptune [/gas giant] mass outflow rate approaches $10^{12}$ g sec$^{-1}$, i.e., comparable to the strongest outflows expected from close-in planets around late main-sequence stars. Whereas the gas giant remains virtually unaffected from an evolutionary standpoint, atmospheric escape may have sizable effects for the sub-Neptune, depending on its dynamical history, e.g., assuming that the hydrogen-helium envelope makes up 1 [/4] per cent of the planet mass, the entire envelope would be evaporated away so long as the planet reaches 0.02 AU within the first 230 [/130] Myr of the WD formation. We discuss how these results can be generalized to eccentric orbits with effective semi-major axis $a'=a/(1-e^2)^{1/4}$, which receive the same orbit-averaged irradiation. Extended to a much broader parameter space, this approach can be exploited to model the expected demographics of WD planets as a function of their initial mass, composition and migration history, as well as their potential for habitability.
\end{abstract}


\section{Introduction}

The systematic search for Earth-like biosignatures in habitable zone planets around G-type stars is currently prohibitive for the next two decades. In spite of their high activity levels, nearby low-mass stars have thus emerged as favorable targets. More recently, the discovery space has expanded to include planets around dead stars, such as white dwarfs (WDs). To date, a handful of exoplanet candidates have been discovered around WDs, including one within $\sim$$0.02$ AU (\citealt{Vanderburg2020}; see also \citealt{Veras2021} and references therein). 
Additionally, between 25-20\% of WDs exhibit atmospheric metal contamination, likely from in-falling planetary debris \citep{farihi09,klein10,zuckerman10,koester14}. 

With their small radii and low luminosities, WDs are challenging targets for planet searches with traditional techniques. However, the large ratio of planet to WD radius makes any such system a relatively inexpensive target for follow-up transit spectroscopy. Indeed, \citet{kaltenneger20} articulated the notion of a ``window of opportunity" for detecting biosignatures around WD habitable planets, should they exist. More recently, \cite{limbach22} demonstrated how the near-featureless spectra of nearby WDs can be exploited to detect an infrared excess from cold Jovian exoplanets with modest exposures.
Systematic searches for cold Jovian and (transiting) terrestrial planets around nearby WDs are currently under way with JWST and TESS, respectively.  

At the same time, a robust theoretical framework for the formation and evolution of WD planets is still lacking. First generation systems beyond $\sim$5 AU may survive the red giant phase of the host star; some planets (or perhaps their moons) could then migrate inward via planet-planet scattering \citep{veras2016,chamandy21}, Kozai-Lidov oscillations \citep{stephan17}, or common envelope evolution \citep{lagos21}. Alternatively, second generation systems could form after the WD itself \citep{debes2002,zuckerman10,lieshout18}, although the ability of any such planet to actually develop and/or retain the conditions for life under these circumstances is far from obvious \citep{agol11,fossati12,barnes13,kozakis20}. 

Absorption of high-energy stellar radiation in the upper atmosphere can significantly affect the evolution of close-in exoplanets. X-ray (5-100 \r{A}) and EUV (100-912 \r{A}) photons--collectively termed XUV--are absorbed at high altitude, in the lower density parts of the atmosphere, where they  {dissociate and/or ionize both molecules and atoms}. The resulting free electrons heat the gas collisionally. These processes inflate the gas and drive hydrodynamic outflows that can potentially remove the primordial atmospheric envelope of hydrogen and helium, a process known as XUV-driven atmospheric escape \citep{Vidal-Madjar2003,Yelle2004,Tian2005,Koskinen2014}. 

Along with core-powered mass loss \citep{ginzburg18, gupta19},  
{photoionization-driven} atmospheric escape is thought to play a significant role in shaping the observed properties of the Kepler planet population \citep{fulton17}, likely contributing to carving the observed radius valley by stripping sub-Neptune-sized planets of their primordial envelopes, and turning them into rocky cores (e.g., \citealt{owenwu13,owenwu17, Rogers2021, Affolter2023, Owen2023}). \\

Because of the vastly different regime of parameter space realized by WDs, in particular the different XUV/bolometric ratios and radii compared to FGK-type stars, which dominate the Kepler sample, the photoionizing radiation arising from young WDs is expected to affect the atmosphere of close-in planets in fundamentally different ways. Depending on when and where WD planets form and/or migrate inward, WD-driven outflows may yield different exoplanet properties compared to those of Kepler stars. The WDs can drive far more efficient mass loss rates (compared to young stars) at very early stages, and much less (to virtually none) after the first few hundred million years.  This sequence of events affects both the expected planet radius distribution and their potential for habitability. Shedding any previously accreted hydrogen-helium envelopes is also a prerequisite for habitability, as the presence of a massive, light element envelope dramatically increases surface temperature and pressure, making it impossible for liquid water to exist, regardless of whether the planet resides in the habitable zone. \\

Beyond a simple energy-limited approach (e.g.,  \citealt{Watson1981}; see also below), the range of parameters pertaining to WD-driven atmospheric outflows is completely unexplored in the literature (see \citealt{schreiber19}, e.g., on the ultimate fate of the Solar System gas giants using energy-limited atmospheric escape). Addressing these questions quantitatively requires an investigation of the physics of XUV-driven photoevaporation for the specific case of a WD spectrum. This Paper takes the first step in this direction by considering the idealized case of a sub-Neptune-sized planet and a gas giant in a circular orbit with radius $a=$0.02 AU around a rapidly cooling WD.  We also show how these results can be generalized to planets on eccentric orbits subject to the same (orbit-averaged) irradiation. 
\begin{figure*}[t!]
\centering{
    \includegraphics[width=0.9\columnwidth]{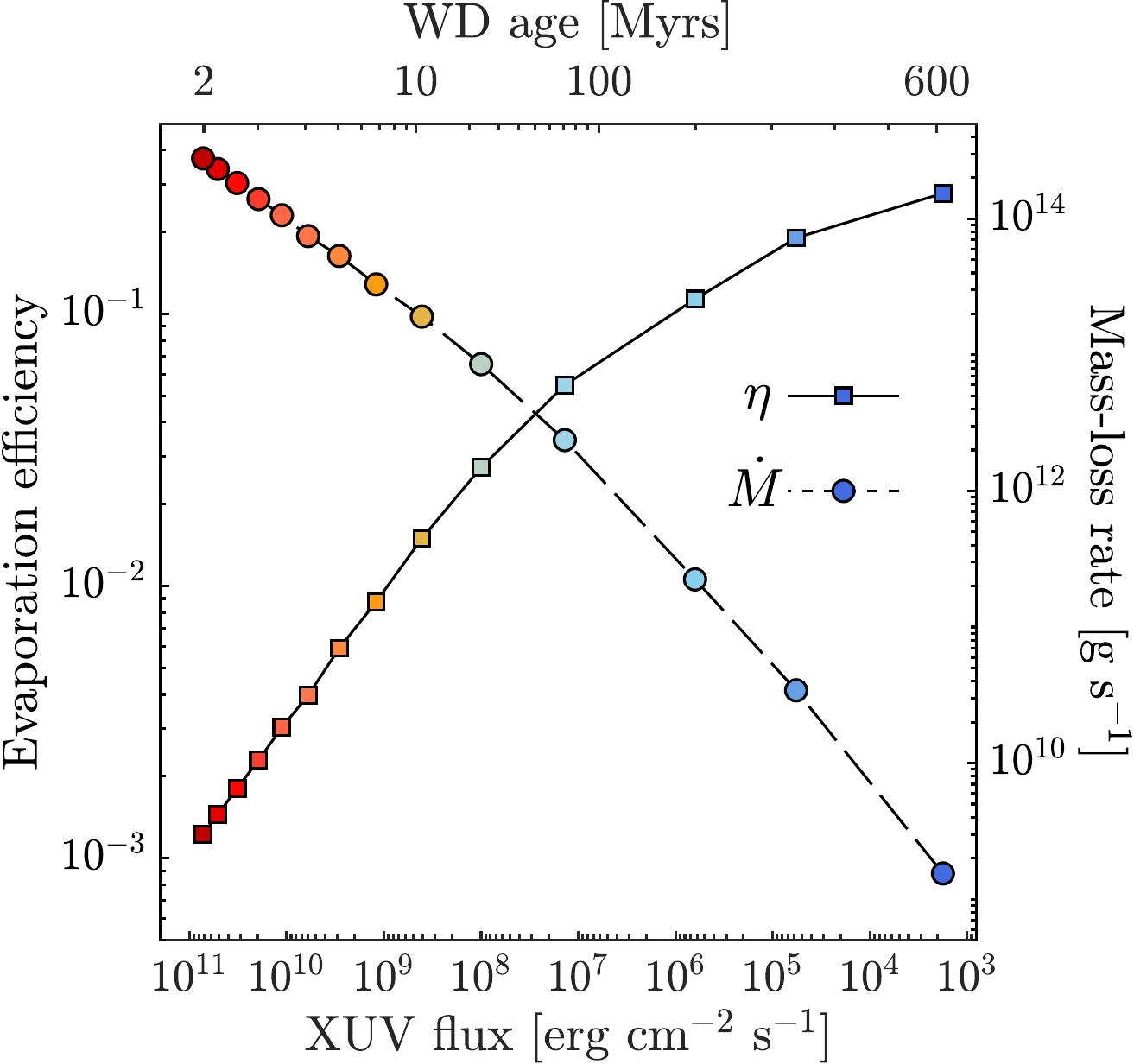}\hspace{1cm}
    \includegraphics[width = 0.89\columnwidth]{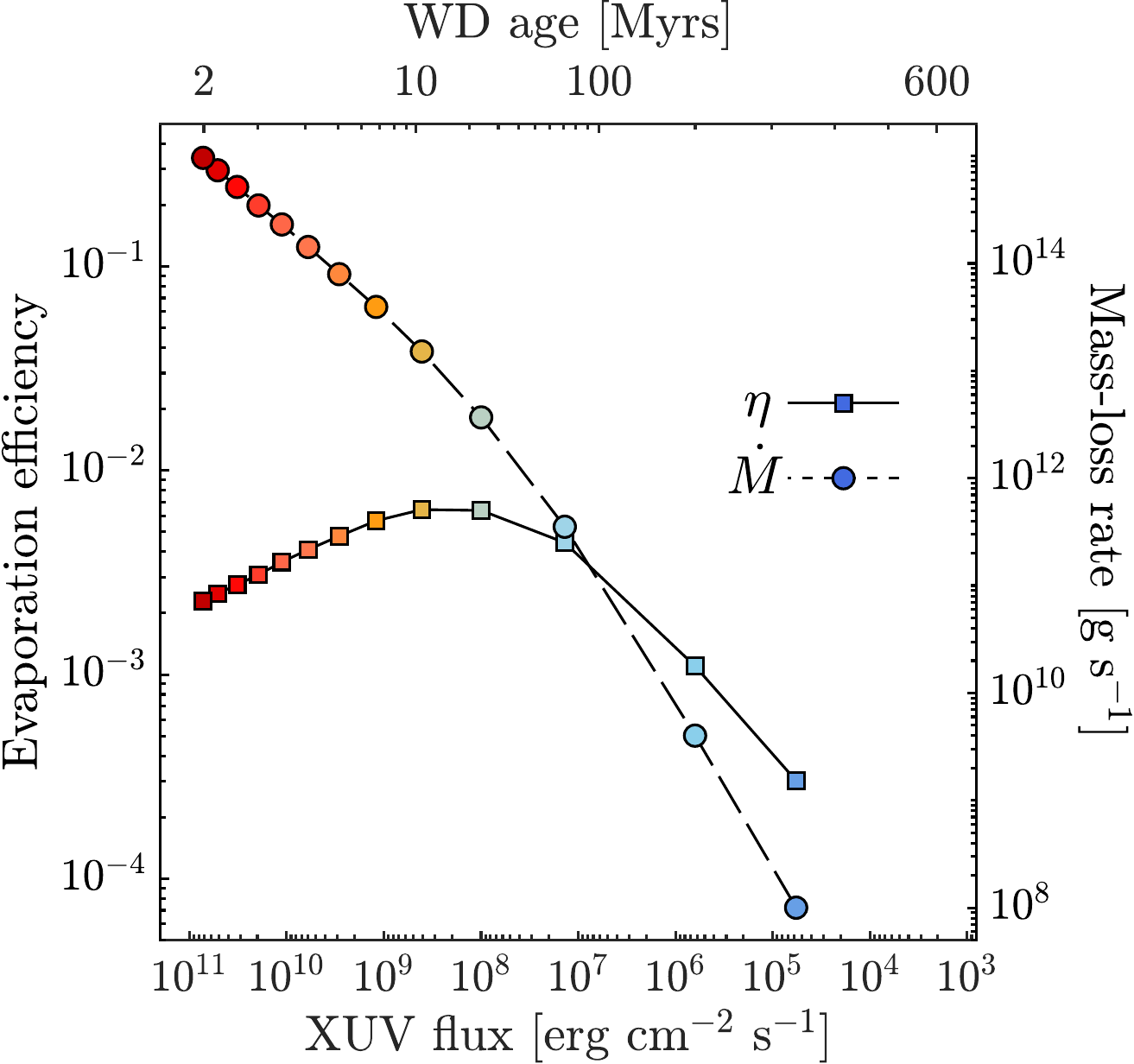}
     \ \\[.5cm]
    \includegraphics[width = 0.85\columnwidth]{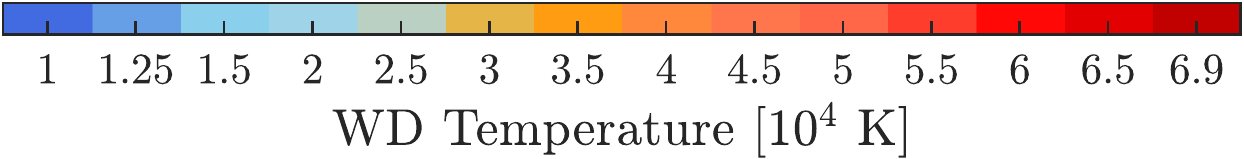}
}
\caption{Left: evaporation efficiency (squares, left y-axis) and atmospheric mass loss rate (circles, right y-axis) as a function of XUV irradiation (bottom x-axis) and or WD age (top x-axis) for the case of a sub-Neptune-sized planet (8 $M_{\Earth}$, 2.7 $R_{\Earth}$) orbiting a 0.61 \msun\ WD at 0.02 AU. The symbol colors trace the WD cooling history (top axis), with $T_{\rm WD}$ varying between 69,000\,K (deepest red) and 5,000\,K (deepest blue). Right: same as left, but for the case of a gas giant (1.24 $M_{\rm J}$ and 1.19 $R_{\rm J}$). 
}\label{fig:eta}
\end{figure*}
\section{From stellar to WD-driven atmospheric escape}
Assuming that all of the stellar XUV flux that is absorbed by a planet atmosphere is converted into expansion work, the instantaneous ion mass loss rate can be expected to scale linearly with the ratio between the incident XUV  {flux at the planet's orbital separation}, $F_{\rm XUV}$, and the average planetary mass density, $\rho_p$ \citep{Erkaev2013}. In practice, numerical work has shown conclusively that the validity of this approximation (known as energy-limited escape) is relatively narrow, as radiative losses can seldom be neglected \cite[e.g.][]{MurrayClay2009,Owen2012,Erkaev2016,Salz2016a,Salz2016b,ates2}. This complication warrants a numerical treatment of the problem. 

\subsection{Numerical Approach}

Our starting point is the publicly available code ATmospheric EScape\footnote{ \citet{ates1}, ATES, v2.0, \href{https://doi.org/10.5281/zenodo.10775751}{doi:10.5281/zenodo.10775751}, as developed on Github: \url{https://github.com/AndreaCaldiroli/ATES-Code} } (ATES). 
For a given planetary system architecture (planet mass, radius, equilibrium temperature, average orbital separation,  {stellar mass} and input stellar spectrum) ATES computes the temperature, density, velocity, and ionization fraction profiles of a highly irradiated H-He atmosphere, and calculates the instantaneous, steady-state mass-loss rate. 
This computation is done by solving the mono-dimensional Euler, mass, and energy conservation equations in radial coordinates through a finite-volume scheme;  {the expression for the gravitational potential also accounts for the Roche potential \citep{Erkaev2007}, which may significantly affect the inferred mass loss rates}. The hydrodynamics module is paired with a photoionization equilibrium solver that includes cooling via bremsstrahlung, recombination, and collisional excitation and ionization, whilst also accounting for advection of the different ion species.  ATES has been validated through a systematic comparison against The PLUTO-CLOUDY Interface (TPCI; \citealt{Salz2015}), a publicly available interface between the magneto-hydrodynamics code PLUTO \citep{pluto} and the plasma simulation and spectral synthesis code CLOUDY \citep{cloudy} (the photoionization equilibrium equations are solved ``on the fly'' in ATES, which results in a factor 80 speed-up of the computing time compared to TPCI). 
In its current, publicly available version, ATES  {recovers stable, steady-state solutions for $\phi_p \simlt 12.9 + 0.17\log F_{\rm XUV}$ (in cgs units), where $\phi_p$ is the planet's gravitational potential energy.} The code is well suited to model  {photoionization-driven} atmospheric escape from planets (i) with radii in excess of $\simgt 1.7 R_{\Earth}$ (i.e., sub-Neptune-sized, or larger), for which a sizable hydrogen-helium envelope may be retained \citep{lopezfortney}, and (ii) at distances within $\sim$0.5 A.U., beyond which atmospheric mass loss is typically negligible (Jeans escape limit), and within which the upper thermosphere comprises only {atomic} (rather than molecular) hydrogen. 

Hereafter, we define the evaporation efficiency {\footnote{Note that this differs from the heating efficiency defined by, e.g., \cite{shematovic14}.}} $\eta$ as the ratio between the mass outflow rate inferred numerically ($\dot{M}$) and the well-known energy-limited approximation for the case where {all} of the absorbed XUV flux is converted into abiabatic expansion: 
\begin{equation}
\eta=\dot M\left(\frac{3F_{\rm XUV}}{4G K \rho_p}\right)^{-1},
\end{equation}
where $K$ accounts for the host star gravitational pull \citep{Erkaev2007}. 
The evaporation efficiency is set primarily by the planet's gravity. Specifically, atmospheric mass loss is always far less efficient ($\eta\ll 1$) than the energy-limited benchmark for gas giants (more specifically, for high-gravity planets, with $\phi_{p} \simgt 14 \times 10^{12}$ erg g$^{-1}$).  {As demonstrated in Section 3 of \cite{ates2}, this inefficiency arises because, after accounting for photo-ionization losses, the mean energy that remains available to heat the ions in the atmosphere is lower than the ions' gravitational binding energy. }
For a fixed  {value of $\phi_{p}$}, the evaporation efficiency of high-gravity planets increases with XUV irradiation, as the fractional contribution of adiabatic cooling increases with flux at a faster pace than Ly$\alpha$ losses, which are the dominant radiative cooling channel.  

\begin{figure*}
    \centering
    \includegraphics[width = 0.75\columnwidth]{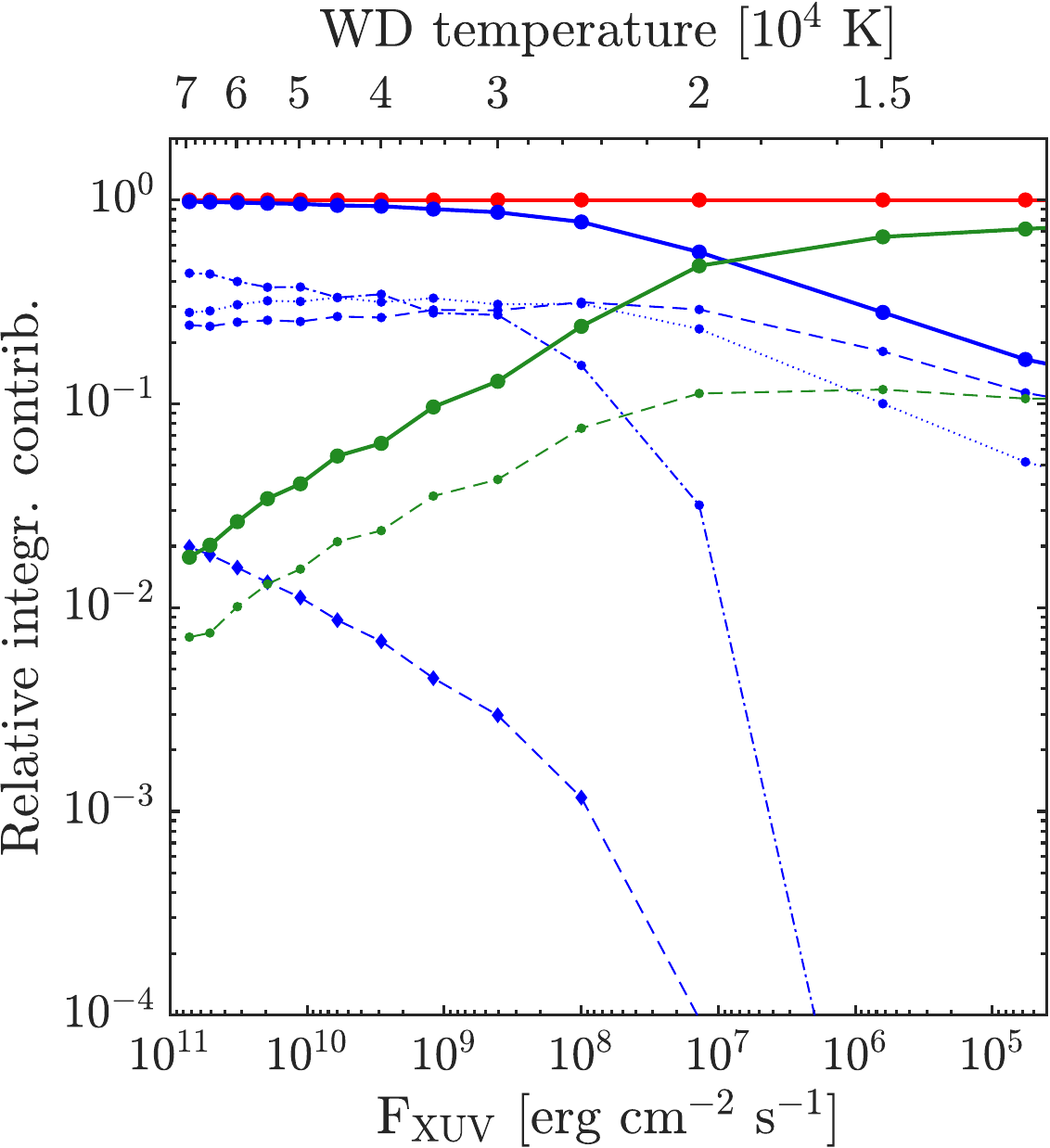}
    \hspace{1cm}
    \includegraphics[width = 0.75\columnwidth]{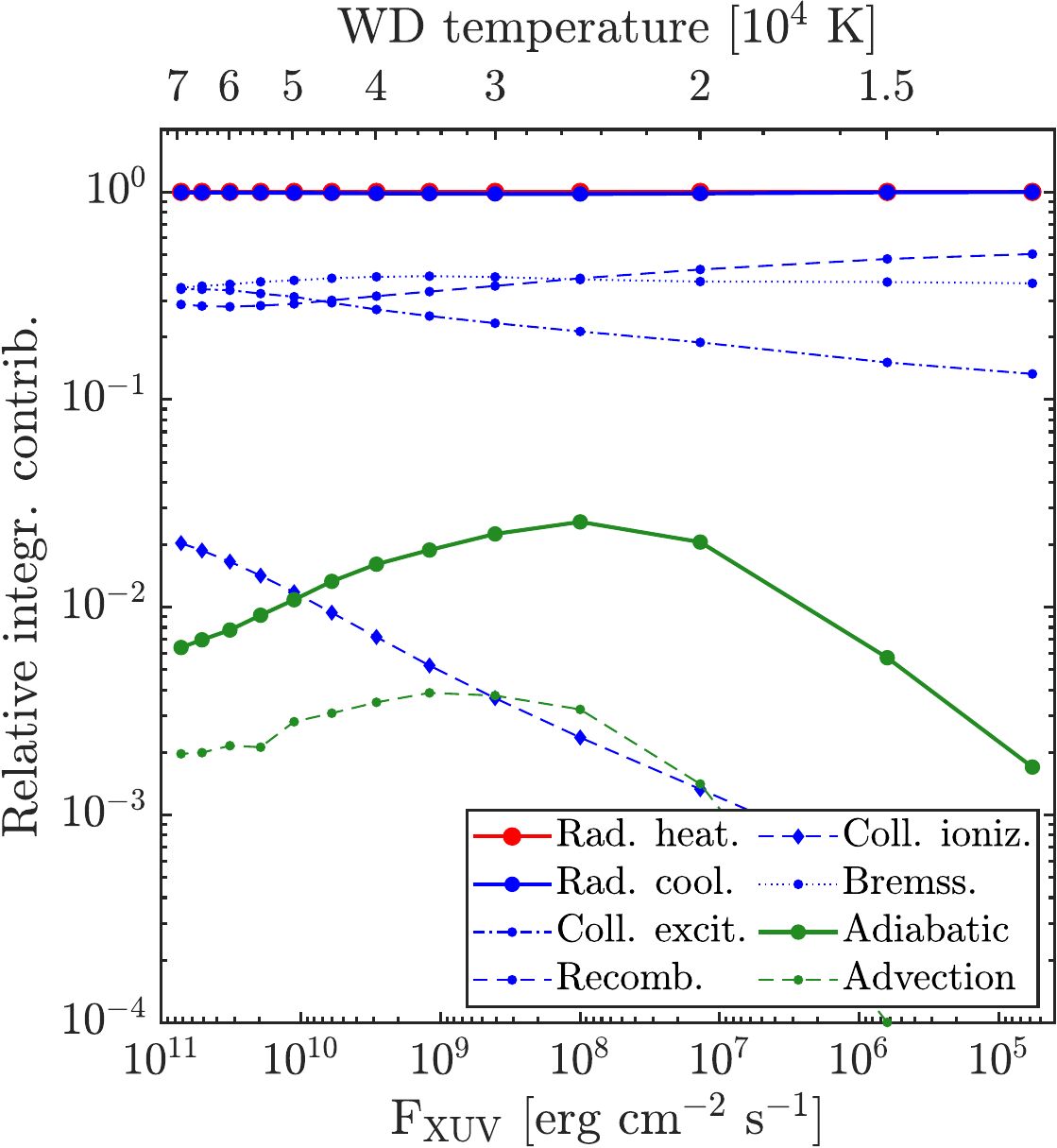}
    \caption{ {Fractional contributions of the {\it total} radiative cooling rate (solid blue line) and abiabatic cooling rate (solid green line) with respect to the heating rate (solid red line)}, shown as a function of the XUV irradiation/WD temperature for the sub-Neptune (left) and gas giant (right). For the  {latter}, the non-monotonic behavior of the abiabatic cooling is responsible for the non monotonic trend of the evaporation efficiency shown in the right panel of Figure \ref{fig:eta}. }
    \label{fig:cooling}
\end{figure*}
Conversely, for lower gravity planets (i.e., Neptune-sized and smaller, with $\phi_p\simlt 8 \times 10^{12}$ erg g$^{-1}$) the mass loss rate is largely independent of $\phi_p$, and is regulated primarily by $F_{\rm XUV}$. These planets can be expected to exhibit energy-limited outflows only below a critical irradiation level ($\eta\simeq 1$ for $F_{\rm XUV} \simlt 10^5~\rm erg\, cm^{-2}\, s^{-1}$). As the irradiation increases, however, the evaporation efficiency decreases, by up to an order of magnitude (even though the mass loss rate increases). This trend is due to a progressive increase in the fractional contribution of advective cooling, followed by a sharp surge in Ly$\alpha$ cooling, with respect to purely adiabatic losses. 

This framework enables a reliable, physically motivated prediction of the  {photoionization-driven mass outflow rate} for a given known or theoretical planetary system  {in close orbit around a late main sequence star (whilst foregoing other mechanisms for driving atmospheric outflows, such as core-powered mass loss and/or boil-off). We note that the above results are fairly insensitive to the assumed X-ray and EUV spectral shape, so long as the combination of shape {\it and} normalization preserve the total number of photons per unit time in each band.} 

We aim to replicate a similar investigation for the parameter space pertaining to planets orbiting a hot WD. 
To model a cooling WD, we make use of BaSTI\footnote{\url{http://basti-iac.oa-abruzzo.inaf.it/wdmodels.html}} \citep{salaris22} stellar evolution models. These consider Carbon-Oxygen WDs with masses of 0.54, 0.61, 0.68, 0.77, 0.87, 1.0, and 1.1 \msun, a range of metallicities\footnote{The Ne mass fraction is a good approximation of the metal mass fraction Z of the progenitor star; BaSTI models are available for Z=0.04, 0.03, 0.017, 0.01, 0.006, and without Ne.}, hydrogen and helium or pure helium atmospheres.
Throughout, we consider a 0.61 \msun\ WD with Z=0.017, and mixed atmosphere. 

 {Unlike for late main sequence stars}, for high temperature/young ages the bulk of the WD flux is emitted in the EUV portion of the spectrum, that is, the energy range that is primarily responsible for driving hydrodynamic outflows. Such extreme ionizing fluxes can be expected to have dramatic consequences on the outflow properties. The ensuing numerical challenges are two-fold. Firstly, we can expect the development of an extremely sharp ionization front close to the inner boundary of the computational domain. Under-resolved ionization fronts translate into sharp discontinuities in the heating rate, which cannot be handled by the code hydrodynamic solver. In addition, non-physical shock waves may develop if the initial and boundary conditions are not carefully chosen. Specifically, a moving shock can arise, e.g., if too much energy is deposited into the initial state. These challenges compromise the stability of ATES' numerical scheme. 

To address them, we implement an adaptive grid refinement routine within the domain of the simulation; this modification enables us to handle sharp, stationary ionization fronts. Next, we modify the code so that the atmosphere is irradiated {gradually} over the first few thousands time steps; this feature avoids the formation of strong shock waves.
\begin{figure*}
 \centering{
    \includegraphics[width = 0.925\columnwidth]{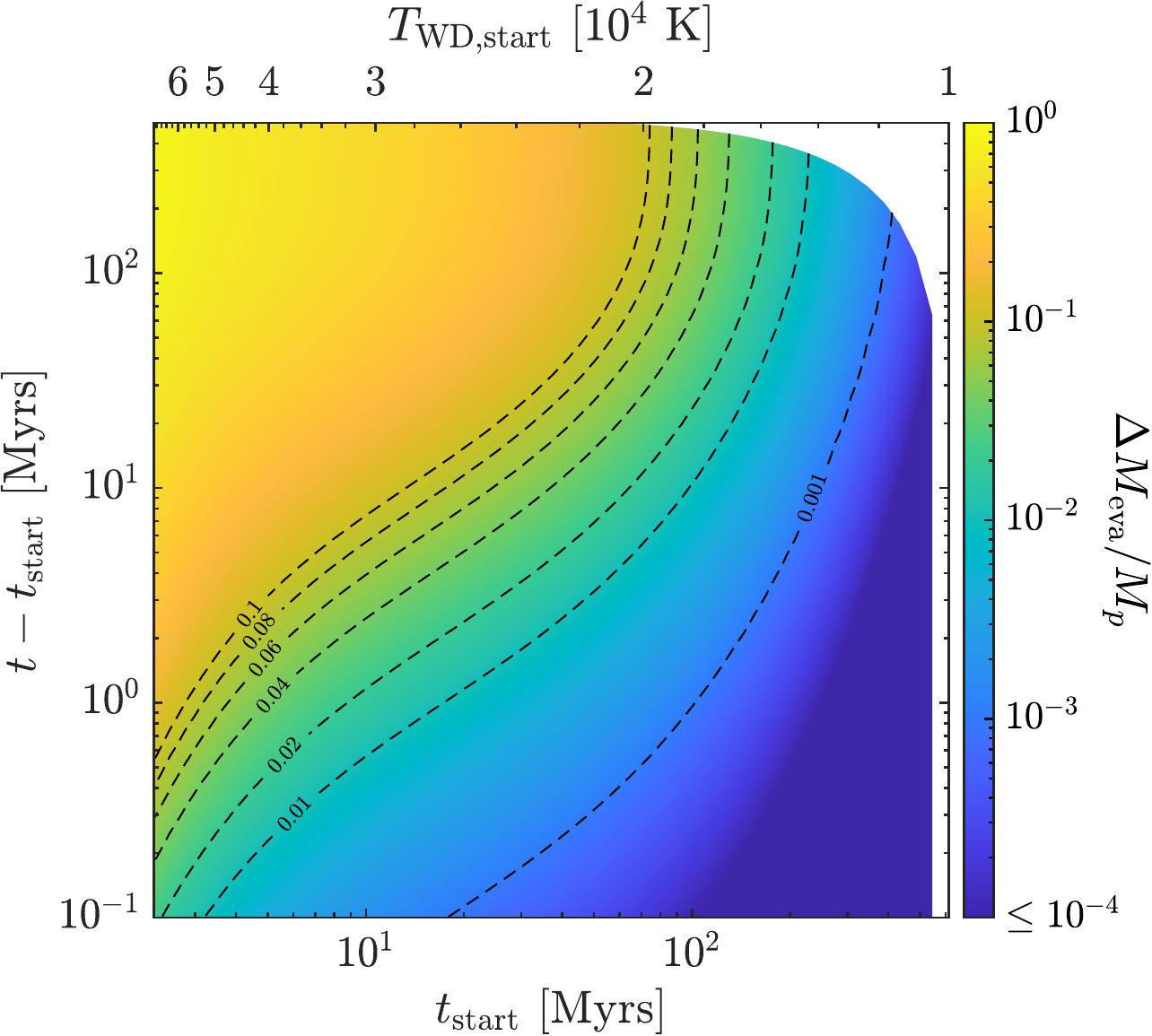}\hspace{1cm}
        \includegraphics[width = 0.925\columnwidth]{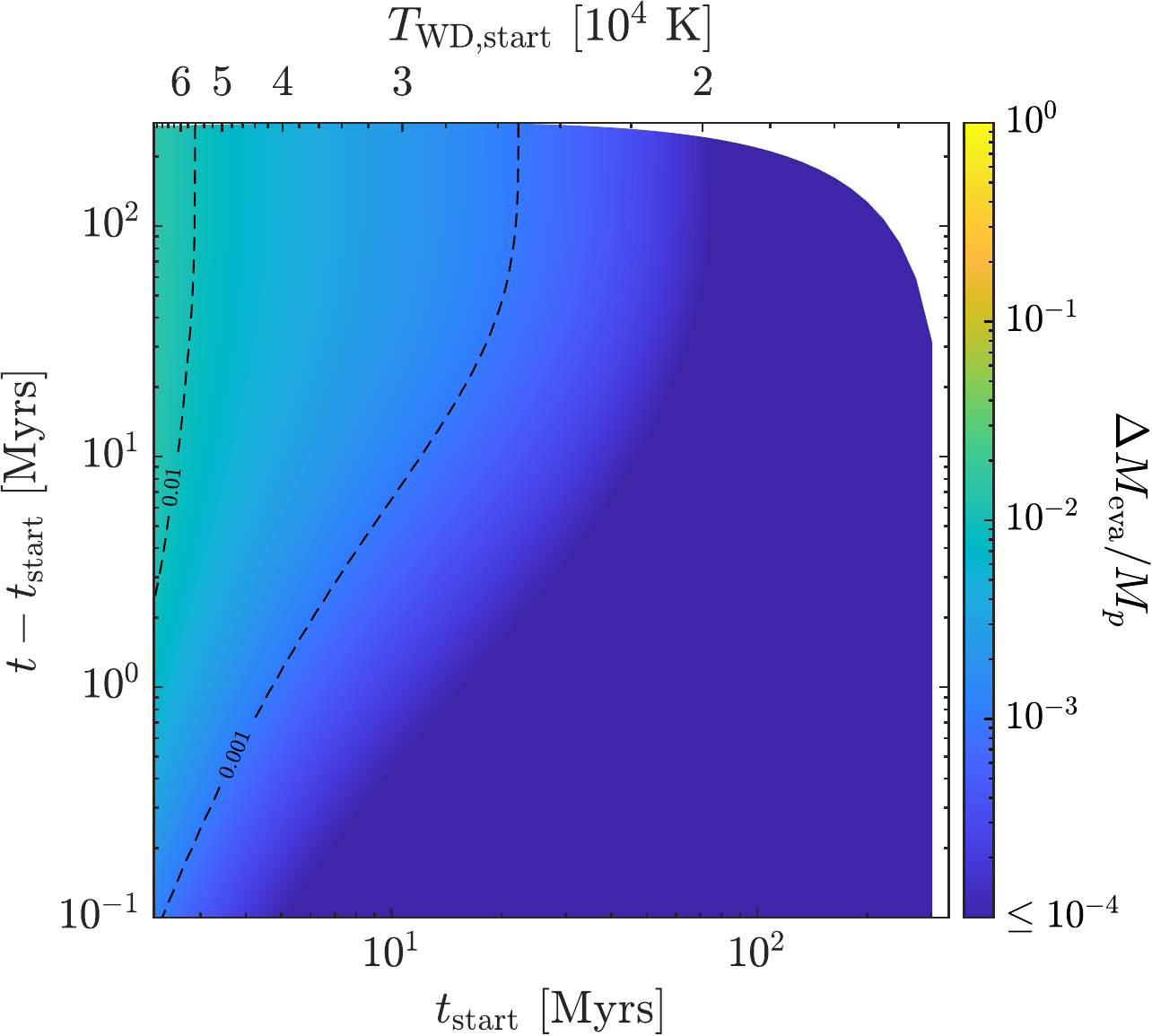}
        }
    \caption{Left: Expected evaporation timescales for the sub-Neptune-sized planet simulated in the left panel of Figure~\ref{fig:eta}. The color bar on the right shows the fractional amount of mass lost to hydrodynamic escape as a function of the planet arrival time at 0.02 AU ($t_{\rm start}$), and of the total residence time at this separation ($t -t_{\rm start}$). The dashed black lines track the locus of the points where the planet would shed 1, 2,..up to 10\% of its mass--in the form of a H-He envelope--via hydrodynamic escape. Right: Same as left, but for the gas giant case in the right panel of Figure~\ref{fig:eta}.
    }
    \label{fig:times}
\end{figure*}

\subsection{Circular Orbit Results}

The modifications and additions outlined above enable the code  {reach steady-state solutions up to $F_{\rm XUV} \simeq 10^{11}~\rm erg\, cm^{-2}\, s^{-1}$, which is about five orders of magnitude higher than the highest XUV irradiation considered by \cite{ates2} for input stellar spectra (albeit assuming different planets' equilibrium temperatures). 
For this pilot study we examine} a low-gravity pl anet (8 $M_{\Earth}$ and 2.7 $R_{\Earth}$, i.e., a typical sub-Neptune) and a high-gravity planet (1.24 $M_{\rm J}$ and 1.19 $R_{\rm J}$, i.e., a typical gas giant),  {both on a circular orbit at 0.02 AU of a 0.61 \msun\ WD, where the orbital separation is chosen specifically to match WD 1856+534 b's} \citep{Vanderburg2020}. 

The main results are  {summarized} in Figure \ref{fig:eta}. The initial WD surface temperature is 69,000 K (dark red); the corresponding  {planet equilibrium temperature is $T_{\rm eq}\simeq 3,200$ K}. After 600 Myr, the WD has cooled down to 10,000 K (darkest blue),  {yielding $T_{\rm eq}\simeq 380$ K}. 
For the sub-Neptune case (left panel of Figure \ref{fig:eta}), the mass loss rate exceeds $10^{12}$ g \pers\ for $T_{\rm WD}\simgt$ 18,000 K, i.e., within the first $\sim$100 Myr. Mass outflow rates as high as $10^{9-10}$ g \pers\ are still driven after 600 Myr, when $T_{\rm WD}\simlt 10,000$ K. 
The evaporation efficiency increases steeply, and monotonically, with decreasing $F_{\rm XUV}$, i.e., as the WD cools down. Nevertheless, the outflow is never quite energy-limited (i.e., $\eta \simlt 1$ at all times).  {As shown in the left panel of Figure~\ref{fig:cooling}, this behavior can be understood in terms of the fractional increase in the adiabatic with respect to the (total) radiative cooling rate going towards lower $F_{\rm XUV}$. }\\

The gas giant (right panel of Figure \ref{fig:eta}) shows very low evaporation efficiencies, with $\eta<10^{-2}$ at all WD ages/temperatures. Nevertheless, the ensuing mass outflow rates range between $10^8$ and $10^{15}$ g \pers, with a peak approaching 5 $M_{\Earth}$/Myr. 
The evaporation efficiency increases slightly overt the first $\sim$30 Myr, as $F_{\rm XUV}$ decreases from $10^{11}$ down to $\simeq 10^8$ erg s$^{-1}$ cm$^{-2}$. Only at later times/lower $F_{\rm XUV}$ does $\eta$ decrease with irradiation.
This behavior can be understood again by examining the relative contribution of adiabatic vs. (total) radiative cooling, shown in the right panel Figure~\ref{fig:cooling}. For XUV irradiations below about $\simlt 10^8$ erg s$^{-1}$ cm$^{-2}$, the ratio of abiabatic to radiative cooling decreases with decreasing flux, leading to a decrease in efficiency. 
However, the ratio of abiabatic to radiative cooling increases with decreasing flux between $10^{11}$ and $10^8$ erg s$^{-1}$ cm$^{-2}$, which means that the evaporation efficiency increases with decreasing flux over this range. \\

These results can be converted into approximate atmospheric evaporation timescales.  {Ignoring momentarily how changes to the planet's atmospheric mass fraction affect its radius}, the instantaneous mass loss rates in Figure \ref{fig:eta} can be used to estimate how long it would take for a close-in planet which has retained its light element envelope to shed it away.   
Figure \ref{fig:times} illustrates the fractional amount of mass that the sub-Neptune (left panel) and gas giant (right panel) planet would lose to hydrodynamic escape as a function of (i) the planet ``arrival time" $t_{\rm start}$ at given WD temperature ($T_{\rm WD, start}$), and (ii) its ``residence" time ($t -t_{\rm start}$), respectively.

As for stellar irradiation, hydrodynamic escape will have sizable evolutionary effects only for the case of a close-in (0.02 AU) sub-Neptune-sized planet. Specifically, the dashed black lines in Figure \ref{fig:times} track the locus of the points where the planet would shed 0.1, 1, 2, $\dots$ up to 10\% of its mass (in the form of a hydrogen-helium envelope) via hydrodynamic escape. As an example, assuming that it makes up 1 [/4] per cent of the planet total mass, the entire envelope  {of the sub-Neptune} would be evaporated in less than about 400 Myr so long as the planet arrives at 0.02 AU within the first 230 [/130] Myr of the WD system life. 
Note that, although the plot extends to nearly complete evaporation (i.e., $\sim$80\% $M_p$), the hydrogen-helium envelope of realistic sub-Neptune-sized planets likely does not exceed mass fractions of a few percent \citep{lopez14}. 

Fractional mass losses are far less extreme for the gas giant; this system will only shed close to 1\% of its initial mass if it starts to be exposed to the extreme XUV flux of an extremely young WD, i.e. within the first few Myr. 

In summary, strong irradiation from a very young/hot WDs can be expected to drive significantly greater atmospheric mass outflow rates than a late main sequence stars. 
For the sub-Neptune-sized planet WD-driven atmospheric escape can shed the entire planet's light element envelope only if the planet reaches an orbital separation as close as 0.02 AU within the first few hundreds Mys of the WD formation, which in turn requires a very efficient mechanism for  {shrinking} the orbit to sub-AU separations. The gas giant (at the same orbital separation) remains fairly unaffected, regardless of how fast it reaches such a very close orbit. 
\subsection{Generalization to Eccentric Orbits}
So far, we have explored the idealized case of a sub-Neptune and a gas giant on a circular orbit, at 0.02 AU of a rapidly cooling WD, and shown that only in the former case is WD-driven atmospheric escape not negligible. We thus focus on the sub-Neptune case hereafter.  
There are several reasons why extending these calculations to the case of non-zero eccentricity might be highly desirable, starting with the fact that numerical work indicates that scattering events, leading to highly eccentric orbits, are likely the most efficient mechanism to drive this kind of planets within sub-AU periastron distances of young WDs \citep{veras2016}.

Whereas we defer a full exploration of the parameter space to a follow-up work, below we examine how the results presented above can be extended to orbits of arbitrary eccentricity $e$$\neq$0. To do so, we wish to derive an analytical expression that characterizes eccentric orbits with the {same orbit-averaged irradiation} as the circular orbit case explored above (i.e., $e=0$ and $a=0.02$ AU). 
We start by deriving the analytic expression of the flux $\langle F \rangle_T$, averaged over the orbital period $T$, that is received by a planet orbiting a WD (or a star) of luminosity $L$ and mass $M_{\ast}\gg M_p$:
\begin{equation}
    \langle F \rangle_T=\frac{2}{T}\int_{0}^{T/2}{dt\,\frac{L}{4\pi r^2}}=\frac{L}{2\pi T}\int_{r_p}^{r_a}{\frac{dt}{dr}\frac{dr}{r^2}},
    \label{eq:flux1}
\end{equation}
where $r(t)$ is the time-dependent orbital separation, and $r_p$ and $r_a$ are the ($t=0$) pericenter and ($t=T/2$) apocenter distance, respectively. 
The radial component of the orbital velocity $dr/dt$ in Equation~\ref{eq:flux1} can be expressed in terms of the semi-major axis $a$ and of the orbital eccentricity $e$, through the specific energy of the orbit, $\mathcal{E}$: 
\begin{equation}
 \mathcal{E}=\frac{1}{2}\left[\left(\frac{dr}{dt}\right)^2+\left(\frac{\mathcal{L}}{r}\right)^2\right]-\frac{GM_{\ast}}{r}, 
 \label{eq:specen}
\end{equation}
where the tangential component of the orbital velocity is written in terms of the orbit's specific angular momentum $\mathcal{L}$. Since $\mathcal{E}=-GM_{\ast}/(2a)$ and $\mathcal{L}=\sqrt{GM_{\ast}a(1-e^2)}$, Equation~(\ref{eq:specen}) yields: 
\begin{equation}
  \frac{dr}{dt}=\sqrt{\frac{GM_{\ast}}{a}}\sqrt{-1+2\frac{a}{r}-\frac{a^2}{r^2}(1-e^2)}.
  \label{eq:radvel}
\end{equation}
Substituting for $dr/dt$ above in Equation~(\ref{eq:flux1}), with $T=2\pi \sqrt{a^3/(GM_{\ast})}$, gives: 
\begin{equation}
 \langle F \rangle_T =\frac{L}{4\pi^2 a}\int_{a(1-e)}^{a(1+e)}{\frac{dr}{r}\frac{1}{\sqrt{-r^2+2ar- a^2(1-e^2)}}},
    \label{eq:eta2}
\end{equation}
where the integration limits are written as $r_p=a(1-e)$ and $r_a=a(1+e)$. 
Integrating Equation~(\ref{eq:eta2}) by substitution (with $y\equiv a/r$) we find:
\begin{align}
     \langle F \rangle_T & = \frac{L}{4\pi^2 a^2}\int_{1/(1+e)}^{1/(1-e)}{dy\,\frac{1}{\sqrt{-1+2y-y^2(1-e^2)}}} \\
                         & = \frac{L}{4\pi a^2 \sqrt{1-e^2}},
    \label{eq:eta3}
\end{align}
consistent with the result of \cite{adams06}. 
The above derivation shows that, over a single eccentric orbit, a planet experiences an average irradiation which is a factor $(1-e^2)^{-1/2}$ higher compared to the case of a circular orbit {with the same energy}. A different way to express this result is that the same orbit-averaged irradiation that is experienced by a planet on a circular orbit of radius $a$ will be experienced by a planet on a (lower energy) eccentric orbit with semi-major axis $a'=a (1-e^2)^{-1/4}$. 

In order to translate an orbit-averaged flux into an orbit-averaged atmospheric mass loss rate we need to account for the functional dependence of the instantaneous mass loss rate on irradiation. If the scaling is linear (i.e., energy-limited escape), then the orbit-averaged mass loss rate coincides with that of a circular orbit with the same orbit-averaged flux.
The scaling of $\dot{M}$ with irradiation becomes shallower at higher irradiation, as a progressively higher fraction of the absorbed WD radiation is used up in radiative processes as opposed to adiabatic expansion. In the limiting case where $\dot{M}$ scales as $\propto F_{\rm XUV}^{1/2}$ (i.e., radiation-limited escape; \citealt{MurrayClay2009}), it can be shown analytically (following similar steps as in Equations [1--6]) that the orbit-averaged mass loss rate is a factor $(1-e^2)^{1/4}$ lower than the mass loss rate for a circular orbit with the same orbit-averaged flux.   

The intermediate regime between energy- and radiation-limited warrants a numerical treatment. For three specific choices of eccentricity and semi-major axes (i.e., $e=0.9,0.95, 0.98$, and $a'=0.03, 0.036, 0.044$ AU, respectively\footnote{The corresponding pericenter [/apocenter] distances are as small [/large] as $r_p\simeq (0.003,0.0018,0.0009)$\,AU [$r_a\simeq(0.057,0.07,0.087)$ AU], respectively.}, such that the orbit-averaged flux equals that of a circular $a=0.02$ AU orbit), we calculate $r(t)$ by integrating Equation (\ref{eq:radvel}). From this, we derive the time-dependent irradiation $F_{\rm XUV}(t)$ along the orbit. To obtain the time-dependent mass loss rate, and its average value, we interpolate the $\dot{M}(F_{\rm XUV})$ results in the left panel of Figure \ref{fig:eta} through a cubic spline function. 
This confirms that progressively increasing the eccentricity (while reducing the orbit's energy) reduces the average mass loss rate along a single orbit, albeit the difference is a factor of about 2 at most. As expected, the differences are more pronounced at high irradiation. This finding is consistent with the analytic results above, whereby the circular to eccentric mass loss rate ratio is $\simlt (1-e^2)^{-1/4}$. 

We emphasize that the above estimates can only be taken as indicative, as they ignore the fact that the size of the system's Roche lobe varies with eccentricity. 

\section{Summary and Future Work}
This work presents the first numerical investigation of planetary atmospheric escape driven by strong irradiation from a hot/young WD. We have examined the specific case of a sub-Neptune planet and a gas giant on a 0.02 AU circular orbit around a rapidly cooling 0.6 \msun\ WD, and showed how the ensuing mass loss rates and evaporation time-scales can be extended to lower energy, eccentric orbits with the same orbit-averaged irradiation. In a forthcoming paper (Caldiroli et al., in prep.) we plan to vastly expand the parameter space, and to carry out proper evolutionary models, assuming different core compositions, envelope mass fractions, and using theoretical mass-radius relations to update the planetary parameters at each time step. 

As shown by \cite{adams11} and \cite{owen_adams14}, any appreciable planetary magnetic field (e.g., 1 G or stronger) is likely to exert a greater magnetic pressure than the ram pressure of the hydrodynamic outflow over a large portion of the parameter space; this acts to limit the net mass outflow rate. Extending this kind of investigation to close-in planets around very young/hot WDs is necessary to attain a more realistic estimate of their cumulative mass loss.\\

Notwithstanding the limitations of this approach (we do not know when/where WD planets formed), the results can be reversed to {place constraints on the residence time of a given planet at a given orbital distance}, and to argue in favor or against the need for very efficient migration. 
The model presented here can be generalized to new WD planet candidates, or used to derive a theoretical maximum (present-day) WD planet mass as function of the parent mass for different evolutionary histories. 
For the specific case of planets within the outer edge of the ``continuously habitable zone" \citep{agol11}, it can be used to estimate the relevant timescales over which any primordial/residual hydrogen-helium envelope can be photo-evaporated away--a prerequisite for habitability. 
Whether or not WDs indeed host a sizable planet population, the observational community has already chosen to invest massively in this opportunity. 
As with M dwarf hosts, virtually every candidate planet is going to be follow-up for spectroscopy. Parallel efforts on the modeling side are necessary to ensure a physically motivated interpretation of WD planets, inform future observational campaigns, and attempt to predict/model their demographics.\\

This work was supported by NASA through Grant No. 80NSSC24K01254 (EG). FCA is supported in part by the Leinweber Center for Theoretical Physics at the University of Michigan. We are grateful to Massimo Dotti and Fabio Rigamonti for useful suggestions on the derivation of the orbit-averaged flux. 

\bibliographystyle{aasjournal}

\begin{thebibliography}{}
\expandafter\ifx\csname natexlab\endcsname\relax\def\natexlab#1{#1}\fi
\providecommand{\url}[1]{\href{#1}{#1}}

\bibitem[{{Adams}(2011)}]{adams11}
{Adams}, F.~C. 2011, \apj, 730, 27

\bibitem[{{Adams} \& {Laughlin}(2006)}]{adams06}
{Adams}, F.~C., \& {Laughlin}, G. 2006, \apj, 649, 992

\bibitem[{{Affolter} {et~al.}(2023){Affolter}, {Mordasini}, {Oza},
  {Kubyshkina}, \& {Fossati}}]{Affolter2023}
{Affolter}, L., {Mordasini}, C., {Oza}, A.~V., {Kubyshkina}, D., \& {Fossati},
  L. 2023, \aap, 676, A119

\bibitem[{{Agol}(2011)}]{agol11}
{Agol}, E. 2011, \apjl, 731, L31

\bibitem[{{Barnes} \& {Heller}(2013)}]{barnes13}
{Barnes}, R., \& {Heller}, R. 2013, Astrobiology, 13, 279

\bibitem[{{Caldiroli} {et~al.}(2021){Caldiroli}, {Haardt}, {Gallo}, {Spinelli},
  {Malsky}, \& {Rauscher}}]{ates1}
{Caldiroli}, A., {Haardt}, F., {Gallo}, E., {et~al.} 2021, \aap, 655, A30.
\newblock \url{https://doi.org/10.1051/0004-6361/202141497}

\bibitem[{{Caldiroli} {et~al.}(2022){Caldiroli}, {Haardt}, {Gallo}, {Spinelli},
  {Malsky}, \& {Rauscher}}]{ates2}
---. 2022, \aap, 663, A122

\bibitem[{{Chamandy} {et~al.}(2021){Chamandy}, {Blackman}, {Nordhaus}, \&
  {Wilson}}]{chamandy21}
{Chamandy}, L., {Blackman}, E.~G., {Nordhaus}, J., \& {Wilson}, E. 2021,
  \mnras, 502, L110

\bibitem[{{Debes} \& {Sigurdsson}(2002)}]{debes2002}
{Debes}, J.~H., \& {Sigurdsson}, S. 2002, \apj, 572, 556

\bibitem[{{Erkaev} {et~al.}(2007){Erkaev}, {Kulikov, Yu. N.}, {Lammer, H.},
  {Selsis, F.}, {Langmayr, D.}, {Jaritz, G. F.}, \& {Biernat, H.
  K.}}]{Erkaev2007}
{Erkaev}, N.~V., {Kulikov, Yu. N.}, {Lammer, H.}, {et~al.} 2007, \aap, 472,
  329.
\newblock \url{https://doi.org/10.1051/0004-6361:20066929}

\bibitem[{{Erkaev} {et~al.}(2016){Erkaev}, {Lammer}, {Odert}, {Kislyakova},
  {Johnstone}, {G{\"u}del}, \& {Khodachenko}}]{Erkaev2016}
{Erkaev}, N.~V., {Lammer}, H., {Odert}, P., {et~al.} 2016, \mnras, 460, 1300

\bibitem[{Erkaev {et~al.}(2013)Erkaev, Lammer, Odert, Kulikov, Kislyakova,
  Khodachenko, Güdel, Hanslmeier, \& Biernat}]{Erkaev2013}
Erkaev, N.~V., Lammer, H., Odert, P., {et~al.} 2013, Astrobiology, 13, 1011,
  pMID: 24251443.
\newblock \url{https://doi.org/10.1089/ast.2012.0957}

\bibitem[{{Farihi} {et~al.}(2009){Farihi}, {Jura}, \& {Zuckerman}}]{farihi09}
{Farihi}, J., {Jura}, M., \& {Zuckerman}, B. 2009, \apj, 694, 805

\bibitem[{{Ferland} {et~al.}(1998){Ferland}, {Korista}, {Verner}, {Ferguson},
  {Kingdon}, \& {Verner}}]{cloudy}
{Ferland}, G.~J., {Korista}, K.~T., {Verner}, D.~A., {et~al.} 1998, \pasp, 110,
  761

\bibitem[{{Fossati} {et~al.}(2012){Fossati}, {Bagnulo}, {Haswell}, {Patel},
  {Busuttil}, {Kowalski}, {Shulyak}, \& {Sterzik}}]{fossati12}
{Fossati}, L., {Bagnulo}, S., {Haswell}, C.~A., {et~al.} 2012, \apjl, 757, L15

\bibitem[{{Fulton} {et~al.}(2017){Fulton}, {Petigura}, {Howard}, {Isaacson},
  {Marcy}, {Cargile}, {Hebb}, {Weiss}, {Johnson}, {Morton}, {Sinukoff},
  {Crossfield}, \& {Hirsch}}]{fulton17}
{Fulton}, B.~J., {Petigura}, E.~A., {Howard}, A.~W., {et~al.} 2017, \aj, 154,
  109

\bibitem[{{Ginzburg} {et~al.}(2018){Ginzburg}, {Schlichting}, \&
  {Sari}}]{ginzburg18}
{Ginzburg}, S., {Schlichting}, H.~E., \& {Sari}, R. 2018, \mnras, 476, 759

\bibitem[{{Gupta} \& {Schlichting}(2019)}]{gupta19}
{Gupta}, A., \& {Schlichting}, H.~E. 2019, \mnras, 487, 24

\bibitem[{{Kaltenegger} {et~al.}(2020){Kaltenegger}, {MacDonald}, {Kozakis},
  {Lewis}, {Mamajek}, {McDowell}, \& {Vanderburg}}]{kaltenneger20}
{Kaltenegger}, L., {MacDonald}, R.~J., {Kozakis}, T., {et~al.} 2020, \apjl,
  901, L1

\bibitem[{{Klein} {et~al.}(2010){Klein}, {Jura}, {Koester}, {Zuckerman}, \&
  {Melis}}]{klein10}
{Klein}, B., {Jura}, M., {Koester}, D., {Zuckerman}, B., \& {Melis}, C. 2010,
  \apj, 709, 950

\bibitem[{{Koester} {et~al.}(2014){Koester}, {G{\"a}nsicke}, \&
  {Farihi}}]{koester14}
{Koester}, D., {G{\"a}nsicke}, B.~T., \& {Farihi}, J. 2014, \aap, 566, A34

\bibitem[{{Koskinen} {et~al.}(2014){Koskinen}, {Lavvas}, {Harris}, \&
  {Yelle}}]{Koskinen2014}
{Koskinen}, T.~T., {Lavvas}, P., {Harris}, M.~J., \& {Yelle}, R.~V. 2014,
  Philosophical Transactions of the Royal Society of London Series A, 372,
  20130089

\bibitem[{{Kozakis} {et~al.}(2020){Kozakis}, {Lin}, \&
  {Kaltenegger}}]{kozakis20}
{Kozakis}, T., {Lin}, Z., \& {Kaltenegger}, L. 2020, \apjl, 894, L6

\bibitem[{{Lagos} {et~al.}(2021){Lagos}, {Schreiber}, {Zorotovic},
  {G{\"a}nsicke}, {Ronco}, \& {Hamers}}]{lagos21}
{Lagos}, F., {Schreiber}, M.~R., {Zorotovic}, M., {et~al.} 2021, \mnras, 501,
  676

\bibitem[{{Limbach} {et~al.}(2022){Limbach}, {Vanderburg}, {Stevenson},
  {Blouin}, {Morley}, {Lustig-Yaeger}, {Soares-Furtado}, \&
  {Janson}}]{limbach22}
{Limbach}, M.~A., {Vanderburg}, A., {Stevenson}, K.~B., {et~al.} 2022, \mnras,
  517, 2622

\bibitem[{{Lopez} \& {Fortney}(2014{\natexlab{a}})}]{lopezfortney}
{Lopez}, E.~D., \& {Fortney}, J.~J. 2014{\natexlab{a}}, \apj, 792, 1

\bibitem[{{Lopez} \& {Fortney}(2014{\natexlab{b}})}]{lopez14}
---. 2014{\natexlab{b}}, \apj, 792, 1

\bibitem[{{Mignone} {et~al.}(2012){Mignone}, {Zanni}, {Tzeferacos}, {van
  Straalen}, {Colella}, \& {Bodo}}]{pluto}
{Mignone}, A., {Zanni}, C., {Tzeferacos}, P., {et~al.} 2012, \apjs, 198, 7

\bibitem[{Murray-Clay {et~al.}(2009)Murray-Clay, Chiang, \&
  Murray}]{MurrayClay2009}
Murray-Clay, R.~A., Chiang, E.~I., \& Murray, N. 2009, \apj, 693, 23.
\newblock \url{https://doi.org/10.1088%2F0004-637x%2F693%2F1%2F23}

\bibitem[{{Owen} \& {Adams}(2014)}]{owen_adams14}
{Owen}, J.~E., \& {Adams}, F.~C. 2014, \mnras, 444, 3761

\bibitem[{Owen \& Jackson(2012)}]{Owen2012}
Owen, J.~E., \& Jackson, A.~P. 2012, \mnras, 425, 2931.
\newblock \url{https://doi.org/10.1111/j.1365-2966.2012.21481.x}

\bibitem[{{Owen} \& {Schlichting}(2023)}]{Owen2023}
{Owen}, J.~E., \& {Schlichting}, H.~E. 2023, arXiv e-prints, arXiv:2308.00020

\bibitem[{{Owen} \& {Wu}(2013)}]{owenwu13}
{Owen}, J.~E., \& {Wu}, Y. 2013, \apj, 775, 105

\bibitem[{{Owen} \& {Wu}(2017)}]{owenwu17}
---. 2017, \apj, 847, 29

\bibitem[{{Rogers} \& {Owen}(2021)}]{Rogers2021}
{Rogers}, J.~G., \& {Owen}, J.~E. 2021, \mnras, 503, 1526

\bibitem[{{Salaris} {et~al.}(2022){Salaris}, {Cassisi}, {Pietrinferni}, \&
  {Hidalgo}}]{salaris22}
{Salaris}, M., {Cassisi}, S., {Pietrinferni}, A., \& {Hidalgo}, S. 2022,
  \mnras, 509, 5197

\bibitem[{{Salz} {et~al.}(2015){Salz}, {Banerjee, R.}, {Mignone, A.},
  {Schneider, P. C.}, {Czesla, S.}, \& {Schmitt, J. H. M. M.}}]{Salz2015}
{Salz}, {Banerjee, R.}, {Mignone, A.}, {et~al.} 2015, A\&A, 576, A21.
\newblock \url{https://doi.org/10.1051/0004-6361/201424330}

\bibitem[{Salz {et~al.}(2016)Salz, Czesla, Schneider, \& Schmitt}]{Salz2016a}
Salz, M., Czesla, S., Schneider, P.~C., \& Schmitt, J. H. M.~M. 2016, \aap,
  586, A75.
\newblock \url{https://doi.org/10.1051/0004-6361/201526109}

\bibitem[{{Salz} {et~al.}(2016){Salz}, {Schneider}, {Czesla}, \&
  {Schmitt}}]{Salz2016b}
{Salz}, M., {Schneider}, P.~C., {Czesla}, S., \& {Schmitt}, J.~H.~M.~M. 2016,
  \aap, 585, L2

\bibitem[{{Schreiber} {et~al.}(2019){Schreiber}, {G{\"a}nsicke}, {Toloza},
  {Hernandez}, \& {Lagos}}]{schreiber19}
{Schreiber}, M.~R., {G{\"a}nsicke}, B.~T., {Toloza}, O., {Hernandez}, M.-S., \&
  {Lagos}, F. 2019, \apjl, 887, L4

\bibitem[{{Shematovich} {et~al.}(2014){Shematovich}, {Ionov}, \&
  {Lammer}}]{shematovic14}
{Shematovich}, V.~I., {Ionov}, D.~E., \& {Lammer}, H. 2014, \aap, 571, A94

\bibitem[{{Stephan} {et~al.}(2017){Stephan}, {Naoz}, \&
  {Zuckerman}}]{stephan17}
{Stephan}, A.~P., {Naoz}, S., \& {Zuckerman}, B. 2017, \apjl, 844, L16

\bibitem[{Tian {et~al.}(2005)Tian, Toon, Pavlov, \& De~Sterck}]{Tian2005}
Tian, F., Toon, O., Pavlov, A., \& De~Sterck, H. 2005, \apj, 621, 1049

\bibitem[{{van Lieshout} {et~al.}(2018){van Lieshout}, {Kral}, {Charnoz},
  {Wyatt}, \& {Shannon}}]{lieshout18}
{van Lieshout}, R., {Kral}, Q., {Charnoz}, S., {Wyatt}, M.~C., \& {Shannon}, A.
  2018, \mnras, 480, 2784

\bibitem[{{Vanderburg} {et~al.}(2020){Vanderburg}, {Rappaport}, {Xu},
  {Crossfield}, {Becker}, {Gary}, {Murgas}, {Blouin}, {Kaye}, {Palle}, {Melis},
  {Morris}, {Kreidberg}, {Gorjian}, {Morley}, {Mann}, {Parviainen}, {Pearce},
  {Newton}, {Carrillo}, {Zuckerman}, {Nelson}, {Zeimann}, {Brown},
  {Tronsgaard}, {Klein}, {Ricker}, {Vanderspek}, {Latham}, {Seager}, {Winn},
  {Jenkins}, {Adams}, {Benneke}, {Berardo}, {Buchhave}, {Caldwell},
  {Christiansen}, {Collins}, {Col{\'o}n}, {Daylan}, {Doty}, {Doyle},
  {Dragomir}, {Dressing}, {Dufour}, {Fukui}, {Glidden}, {Guerrero}, {Guo},
  {Heng}, {Henriksen}, {Huang}, {Kaltenegger}, {Kane}, {Lewis}, {Lissauer},
  {Morales}, {Narita}, {Pepper}, {Rose}, {Smith}, {Stassun}, \&
  {Yu}}]{Vanderburg2020}
{Vanderburg}, A., {Rappaport}, S.~A., {Xu}, S., {et~al.} 2020, \nat, 585, 363

\bibitem[{{Veras}(2016)}]{veras2016}
{Veras}, D. 2016, Royal Society Open Science, 3, 150571

\bibitem[{{Veras}(2021)}]{Veras2021}
---. 2021, in Oxford Research Encyclopedia of Planetary Science, 1

\bibitem[{{Vidal-Madjar} {et~al.}(2003){Vidal-Madjar}, {Lecavelier des Etangs},
  {D{\'e}sert}, {Ballester}, {Ferlet}, {H{\'e}brard}, \&
  {Mayor}}]{Vidal-Madjar2003}
{Vidal-Madjar}, A., {Lecavelier des Etangs}, A., {D{\'e}sert}, J.~M., {et~al.}
  2003, \nat, 422, 143

\bibitem[{Watson {et~al.}(1981)Watson, Donahue, \& Walker}]{Watson1981}
Watson, A., Donahue, T., \& Walker, J. 1981, \icarus, 48, 150 .
\newblock
  \url{http://www.sciencedirect.com/science/article/pii/0019103581901019}

\bibitem[{Yelle(2004)}]{Yelle2004}
Yelle, R.~V. 2004, \icarus, 170, 167 .
\newblock
  \url{http://www.sciencedirect.com/science/article/pii/S0019103504000727}

\bibitem[{{Zuckerman} {et~al.}(2010){Zuckerman}, {Melis}, {Klein}, {Koester},
  \& {Jura}}]{zuckerman10}
{Zuckerman}, B., {Melis}, C., {Klein}, B., {Koester}, D., \& {Jura}, M. 2010,
  \apj, 722, 725

\end{thebibliography}

\end{document}